\documentclass[twocolumn,resetfootnote]{aastex7}

\begin{document}

\title{A CNN--Transformer Denoiser for low-$S/N$ Galaxy Spectra: Stellar Population Recovery in Synthetic Tests}

\author[0000-0003-3474-9047]{Suk Kim}
\affiliation{Korea Astronomy and Space Science Institute, Daejeon 34055, Republic of Korea}
\email[show]{star4citizen@gmail.com}

\author[0000-0003-3451-0925]{Joon Hyeop Lee}
\affiliation{Korea Astronomy and Space Science Institute, Daejeon 34055, Republic of Korea}
\email{jhl@kasi.re.kr}

\author[0000-0002-0041-6490]{Soo-Chang Rey}
\affiliation{Department of Astronomy and Space Science, Chungnam National University, Daejeon 34134, Republic of Korea}
\email{screy@cnu.ac.kr}

\begin{abstract}

Stellar population measurements in integral field unit surveys are often limited by low signal-to-noise ratios ($S/N$) in low-surface-brightness spaxels. Using controlled synthetic experiments, we investigate whether a deep-learning-based denoising can recover stellar population information from such spectra without requiring spatial binning. We introduce the Enhanced U-Net Transformer (EUT), a one-dimensional CNN--Transformer model trained on 90,000 synthetic spectra constructed from MILES simple stellar population (SSP) models following \citet{Lee2023}, with wavelength-dependent noise injected on the fly to emulate SAMI-like data ($S/N\simeq 5$--20, measured in a 4484.77--4573.12~\AA\ continuum window). Utilizing an independent test set of 10,000 spectra, the EUT reduces the full-spectrum root-mean-square (RMS) residual by $\simeq96.5\%$ at $S/N=5$ (and by $\simeq94\%$ at $S/N=20$), achieving recovery rates of $\ge 99.8\%$ (the Pearson correlation coefficient between the noise-free and comparison spectra expressed in percent). In fixed windows around \ion{Ca}{2}~H, H$\delta$, H$\beta$, \ion{Fe}{1}~4383, Mg~b, and Na~D, residuals decrease by $\gtrsim88\%$ while preserving line-profile structure. In downstream analysis with \textsc{pPXF}, we assess parameter recovery using the Pearson correlation coefficient $R_p$ and the RMS scatter: the scatter in recovered mass-weighted age decreases from $\simeq0.41$ to $\simeq0.25$~dex at $S/N=5$ and from $\simeq0.32$ to $\simeq0.22$~dex at $S/N=10$; the corresponding mass-weighted global metallicity, $[\mathrm{M/H}]$, scatter decreases from $\simeq0.45$ to $\simeq0.36$~dex and from $\simeq0.32$ to $\simeq0.28$~dex. At $S/N=20$, denoising yields results consistent with those from the noisy inputs within the synthetic-test uncertainties. These controlled experiments suggest that hybrid CNN--Transformer denoisers can enhance the usable low-surface-brightness area for stellar population studies, although further validation with observed spectra will be needed before practical application.

\end{abstract}

\section{Introduction}

Large-scale integral field unit (IFU) surveys, such as CALIFA \citep{Sanchez2012,GarciaBenito2015}, MaNGA \citep{Bundy2015,Law2016,Westfall2019}, and SAMI \citep{Croom2012}, provide spatially resolved spectroscopy across nearby galaxies. These three-dimensional data cubes facilitate detailed investigations into the internal structure and evolution of galaxies, encompassing phenomena such as inside-out quenching, kinematic disparities between gas and stars, and radial gradients in stellar metallicity and age \citep[e.g.,][]{Perez2013,GonzalezDelgado2014,GonzalezDelgado2015,SanchezBlazquez2014}. Consequently, IFU surveys offer a framework for linking internal galaxy properties to integrated galaxy parameters and environment \citep[e.g.,][]{Sanchez2021,Belfiore2017,Goddard2017}.

A fundamental observational constraint arises from distributing a fixed photon budget across numerous spatial elements, which inevitably reduces the signal-to-noise ratio ($S/N$) per spaxel. This effect is particularly pronounced in regions of low surface brightness, such as the outskirts of massive galaxies or intrinsically diffuse systems. In practice, a substantial fraction of spaxels in current IFU surveys exhibit $S/N \lesssim 5$~\AA$^{-1}$ \citep[e.g.,][]{GarciaBenito2015}. These regions may retain signatures of galaxy assembly and accretion (e.g., ex-situ stellar components; \citealt{Oser2010,Cooper2013}), yet low $S/N$ limits spatially resolved analysis of their stellar population properties.

To mitigate noise, common approaches include spatial binning (e.g., Voronoi binning; \citealt{Cappellari2003}) and spectral smoothing \citep{Worthey1997,Kriek2009}. While adaptive binning increases $S/N$, it can compromise spatial resolution by mixing distinct regions within a bin \citep[e.g.,][]{CidFernandes2013}. Spectral smoothing suppresses high-frequency noise but may also attenuate narrow absorption features and weak line wings that carry age-, metallicity-, and abundance-sensitive information \citep[e.g.,][]{Cappellari2003,Husemann2013,Westfall2019}. Such processing can introduce systematic biases in derived stellar population parameters, effectively replacing random noise with modeling systematics.

Full spectral fitting codes such as \textsc{pPXF} are extensively used for deriving stellar population parameters from integrated spectra \citep{Cappellari2004,Cappellari2017}. However, their performance degrades at low $S/N$, where the age--metallicity degeneracy becomes more severe because Balmer and metal absorption features are measured less precisely, weakening the ability of the fit to distinguish between younger, metal-rich and older, metal-poor populations. Under such conditions, even modest mismatch between the observed spectrum and the adopted template library can lead to noisy or biased solutions \citep[e.g.,][]{Wilkinson2015}. Using synthetic galaxy spectra, \citet{Lee2023} quantified how uncertainties of recovered age and metallicity increase as $S/N$ decreases, demonstrating that conventional full spectral fitting becomes unreliable below survey-dependent $S/N$ thresholds.

Recently, deep-learning-based denoising has been investigated as a potential method for improving low S/N spectra while maintaining the original spatial sampling. Another approach involves using neural networks to directly derive stellar population parameters from noisy spectra, bypassing the need to explicitly reconstruct a denoised spectrum. However, in this study, our primary objective is to evaluate whether spectral denoising can enhance the accuracy of a conventional full-spectrum-fitting analysis in low S/N regime. As a result, we continue to employ \textsc{pPXF} in the subsequent evaluation, allowing us to evaluate the effects of denoising on parameter recovery independently from the fitting method itself. Adopting the same downstream fitting framework is also crucial for making a direct comparison with the benchmark set by J. H. Lee et al. (2023). Convolutional neural networks (CNNs) and related architectures have been applied to denoise astronomical data and model stellar and galaxy spectra \citep[e.g.,][]{Frontera-Pons2017,Scourfield2023,Gebran2025}, yet existing methodologies remain largely confined to pure CNN or autoencoder frameworks. While these models effectively capture local features within a limited receptive field, they may be less effective at modeling long-range correlations across widely separated wavelength regions. This is relevant for galaxy spectra, where physically linked diagnostics may occur far apart in wavelength (e.g., the 4000~\AA\ break and the Ca\,\textsc{ii} triplet; \citealt{Conroy2012}). This necessitates a denoiser capable of capturing both local line structure and long-range spectral dependencies.

In this work, we investigate whether the Enhanced U-Net Transformer (EUT) improves spectra with an input $S/N \simeq 5$ to a level sufficient for stable \textsc{pPXF}-based recovery of stellar population properties in controlled synthetic experiments. By combining a U-Net-style convolutional encoder--decoder with a Transformer bottleneck, the model is designed to maintain detailed absorption-line profiles while learning global spectral dependencies through self-attention. We train the model using an extensive library of synthetic galaxy spectra with realistic wavelength-dependent noise ($S/N \simeq 5$--20), constructed following \citet{Lee2023} and calibrated to resemble the noise characteristics of contemporary IFU data.

Given that ground-truth stellar population parameters are not available for real galaxies, we utilize synthetic data sets in which the underlying spectra and noise properties are fully specified \citep[e.g.,][]{Lee2023}. This controlled setting allows a quantitative assessment of (i) spectral reconstruction fidelity and (ii) downstream recovery of mass-weighted age and metallicity from full spectral fitting, without additional complications from unknown observational systematics.
We therefore assess whether, within this synthetic framework, EUT denoising produces spectra that support stable full spectral fitting and improves the accuracy of recovered stellar population parameters at low $S/N$. Application to real IFU observations will require further work to address instrumental systematics and potential domain mismatch between simulations and data, but the present study establishes a baseline performance in a controlled regime.

The paper is organized as follows: Section~2 describes the synthetic spectra and noise model; Section~3 outlines the EUT architecture and training; Sections~4 and 5 present denoising performance and its impact on \textsc{pPXF}-based stellar-population recovery; Section~6 summarizes.

\section{Data}

To assess the denoising performance in the low-$S/N$ regime, we utilize the library of synthetic galaxy spectra constructed by \citet{Lee2023}\footnote{\url{<https://github.com/star4citizen/denoising_gal_spectra>}}. These spectra reproduce the spectral sampling and noise characteristics of the SAMI Galaxy Survey, allowing for a direct comparison between noisy inputs and ground-truth references.

\subsection{Synthetic Spectra Construction}

We generate 90{,}000 synthetic galaxy spectra by combining MILES simple stellar population (SSP) models linearly \citep{Vazdekis2010}. These spectra form the model-development subset from a total synthetic library of 100{,}000 spectra, with the remaining 10{,}000 spectra set as an independent test set (Section~2.3).
Following \citet{Lee2023}, each spectrum is a random linear combination of five SSP components drawn from the discrete MILES grid spanning ages 0.063--15.85~Gyr (25 values) and six metallicities ($[\mathrm{M/H}]= -1.71, -1.31, -0.71, -0.40, 0.00, +0.22$).

We adopt the Padova+00 isochrones \citep{Girardi2000} and a unimodal initial mass function (IMF) with a logarithmic slope $\Gamma = 1.3$.

These choices impact the synthetic spectra through the adopted stellar evo lutionary tracks and the relative weighting of stars with different masses, thereby affecting the continuum shape and the strengths of absorption lines. In this study, we maintain consistent isochrone and IMF choices across the training, validation, and test datasets to specifically examine the impact of noise and denoising within a controlled synthetic environment. We note, however, that these assumptions remain part of the adopted data model and could impact quantitative performance when extending to other SSP model families or observed spectra.
The five components are assigned random weights normalized to sum to unity, spanning simplified star formation histories. Although these mixtures do not capture the full complexity of real galaxies, \citet{Lee2023} showed that they provide sufficient coverage for testing full-spectrum fitting at low $S/N$. In this work, we exclude dust attenuation and nebular emission to isolate the impact of noise on stellar population recovery.

\subsection{Normalization and Realistic Noise Injection}

Prior to noise injection, all spectra are normalized by the mean flux in a continuum window (4484.77--4573.12~\AA) that is relatively free of strong absorption features. This normalization emphasizes relative spectral-feature strengths (e.g., absorption-line depths) rather than absolute flux scaling. We use the same window to define $S/N$ throughout this work.

We inject wavelength-dependent Gaussian noise into the normalized spectra following \citet{Lee2023}. We use empirical noise curves from the SAMI Galaxy Survey, which encode the wavelength dependence of instrumental throughput and sky background, to set the noise amplitude for each pixel.

For each training epoch and spectrum, we draw a target $S/N$ uniformly from 5 to 20. We define $S/N$ (per wavelength pixel) in the continuum window as the mean continuum level divided by the noise standard deviation in that window. This on-the-fly noise injection exposes the model to a wide range of noise levels and realizations, reducing sensitivity to any particular $S/N$ value or noise instance.

\subsection{Training, Validation, and Test Sets}

From the 90{,}000 noise-free synthetic spectra, we assign 9{,}000 (10\%) to validation and 81{,}000 to training. During training, noisy realizations are generated on the fly as described above. We use a per-GPU batch size of 120 (effective batch size 480 over four GPUs), limited by the 48\,GB memory per GPU.

For testing, we use the independent set of 10{,}000 synthetic spectra provided by \citet{Lee2023}, which does not overlap with our training or validation samples. This enables a direct comparison between our denoising results and the full-spectrum-fitting reliability analysis presented by \citet{Lee2023}.

\begin{figure*}
\centering
\includegraphics[width=\linewidth]{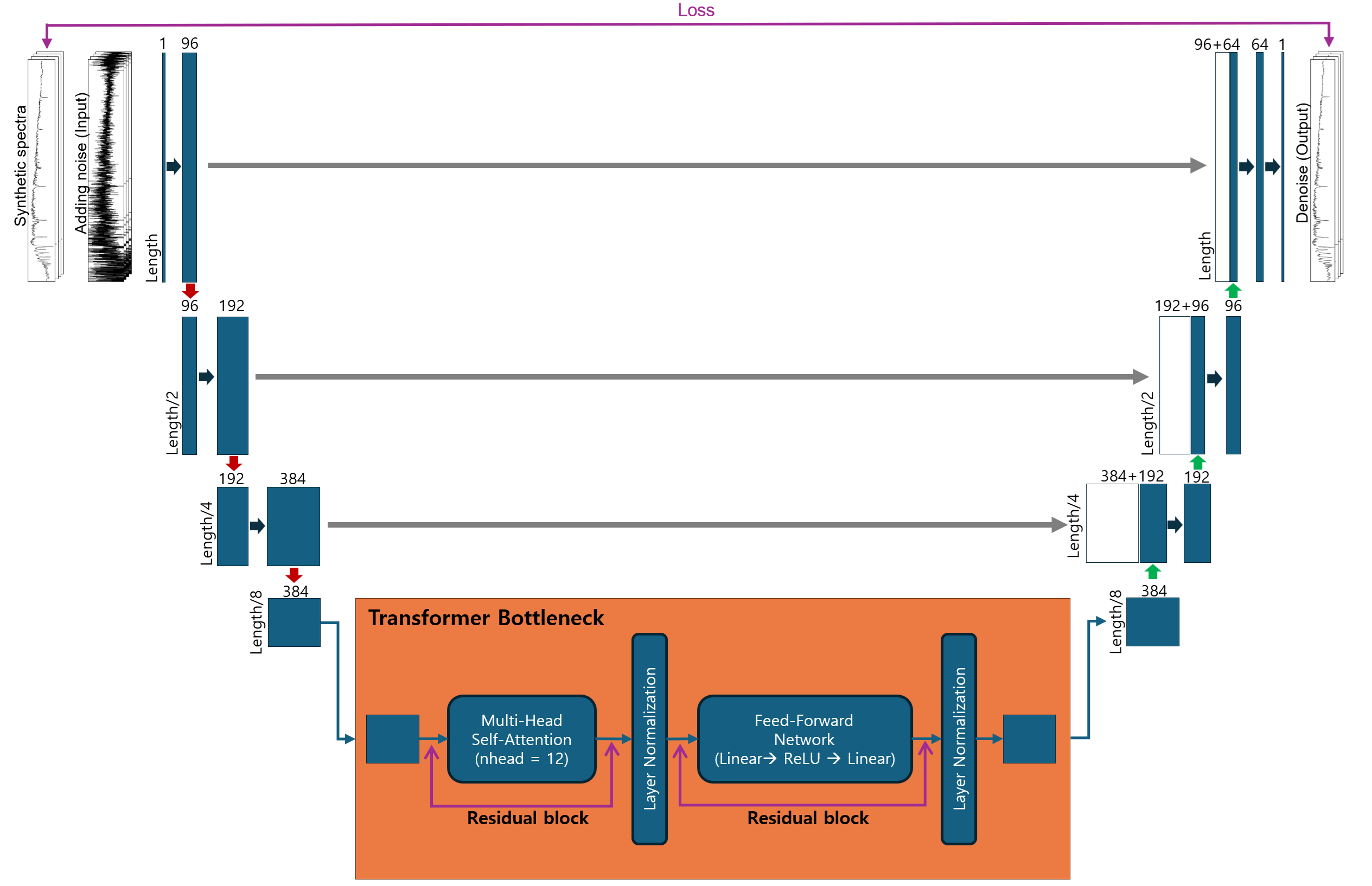}
\caption{
Schematic of the Enhanced U-Net Transformer (EUT) architecture. The network takes a synthetic galaxy spectrum and a noise-added realization as input (left) and predicts a denoised spectrum (right); the loss is evaluated against the corresponding noise-free spectrum. The encoder (blue, left) has three 1D convolutional stages with channel dimensions $1 \rightarrow 96 \rightarrow 192 \rightarrow 384$. Each stage applies a convolution (kernel size $=7$), instance normalization, and ReLU activation, followed by max pooling that halves the sequence length ($L \rightarrow L/2 \rightarrow L/4 \rightarrow L/8$). The compressed feature sequence is passed to a Transformer bottleneck (orange), where fixed sinusoidal positional encodings and $N_{\mathrm{layer}}=8$ Transformer blocks---multi-head self-attention ($n_{\mathrm{head}}=12$) and position-wise feed-forward networks with residual connections and layer normalization---model long-range correlations along the wavelength axis. The decoder (blue, right) upsamples back to the original length using linear interpolation and convolution, with channel dimensions $384 \rightarrow 192 \rightarrow 96 \rightarrow 64$, followed by a final one-channel projection. Gray arrows indicate skip connections that concatenate encoder and decoder feature maps at matched resolutions, helping preserve narrow absorption-line structure in the denoised output.
}
\label{fig:architecture}
\end{figure*}

\section{Denoising Model: Enhanced U-Net Transformer (EUT)}

We present a hybrid deep-learning architecture, the Enhanced U-Net Transformer (EUT), for denoising one-dimensional astronomical spectra. Relative to a standard 1D U-Net \citep{Ronneberger2015}, EUT replaces the bottleneck with a Transformer module \citep{Vaswani2017} and is trained with a composite loss that constrains both local absorption-line structure and the large-scale continuum shape. The model integrates a U-Net-style convolutional neural network (CNN) encoder--decoder for local feature extraction with a Transformer bottleneck that captures long-range spectral dependencies.

\subsection{Network Architecture}

The EUT architecture has three stages: a contracting CNN encoder, a Transformer bottleneck, and an expansive CNN decoder (Figure~\ref{fig:architecture}). The encoder--bottleneck--decoder design, together with skip connections, follows the U-Net framework \citep{Ronneberger2015} and its adaptations for astronomical image and spectral denoising \citep[e.g.,][]{Vojtekova2021,Zhong2025}.

The CNN encoder extracts hierarchical local features from the noisy input spectrum. It includes three blocks, each consisting of a 1D convolution (kernel size $=7$), instance normalization, and a ReLU activation. Each block ends with max pooling (kernel size $=2$), which downsamples the input while increasing the channel dimension ($1 \rightarrow 96 \rightarrow 192 \rightarrow 384$). This multiscale encoding captures features spanning a range of widths, from narrow metal absorption lines to broader molecular bands \citep[e.g.,][]{Zhong2025}.

The Transformer bottleneck mitigates the limited receptive field of standard CNNs when modeling long-range dependencies. The compressed feature map is treated as a sequence of tokens along the wavelength axis and is augmented with fixed sinusoidal positional encodings \citep{Vaswani2017}. The bottleneck contains eight Transformer layers with multi-head self-attention ($n_{\mathrm{head}}=12$; embedding dimension $d_{\mathrm{model}}=384$) and a position-wise feed-forward network. We apply residual connections and layer normalization for training stability. Self-attention enables correlations between widely separated wavelength regions to be learned, improving reconstruction of the large-scale continuum shape \citep[e.g.,][]{Vaswani2017,Leung2024,Zhang2024}.

The CNN decoder reconstructs the denoised spectrum using an expansive path symmetric to the encoder. We upsample the feature maps via linear interpolation followed by convolution, and we concatenate skip-connection features from the encoder at each resolution. These skip connections propagate high-frequency information (e.g., sharp absorption features) from the encoder to the decoder and help preserve fine spectral structure that can be suppressed in deep autoencoders \citep{Ronneberger2015,Vojtekova2021}.

The final decoder block applies a 1D convolution with kernel size 1 and a linear activation to project the features to a single output channel.
We did not determine this architecture by conducting an exhaustive search of hyperparameters or neural architectures. Instead, we chose the final configuration through a limited empirical tuning, focusing on validation performance, training stability, and the memory limitations of the available GPU system as our primary criteria. Consequently, we regard the selected architecture as a practical configuration that worked well in our setup, rather than a formally optimized design. A more thorough investigation of architectural hyperparameters is reserved for future work.

\subsection{Composite Loss Function}

Single-term pixel-space losses (e.g., mean squared error; MSE) may not simultaneously capture narrow spectral features and large-scale continuum trends. We therefore train EUT with a composite objective that combines complementary constraints, following common practice in scientific machine learning \citep[e.g.,][]{Schawinski2017,Jia2019}:
\begin{equation}
L_{\mathrm{total}}
= L_{\mathrm{MSE}}
+ \lambda_{\mathrm{peak}} L_{\mathrm{peak}}
+ \lambda_{\mathrm{fourier}} L_{\mathrm{fourier}}.
\end{equation}

Here $\lambda_{\mathrm{peak}}$ and $\lambda_{\mathrm{fourier}}$ are dimensionless hyperparameters that weight the peak-region penalty (Equation~3) and the low-frequency Fourier-domain loss (Equation~4) relative to the baseline pixel-space MSE, respectively.

We did not use a formal grid search or any quantitative hyperparameter optimization method to determine these weights. Instead, we empirically selected them based on exploratory training runs by monitoring how quickly the total loss decreased in the early training stage and whether the individual loss components showed stable convergence without any single term dominating the optimization process. For all experiments presented here, we use ($\lambda_{\mathrm{peak}}$, $\lambda_{\mathrm{fourier}}$)=(0.1,0.5). By construction, larger $\lambda_{\mathrm{peak}}$ values place more weight on high-flux regions, while larger $\lambda_{\mathrm{fourier}}$ values more strongly enforce low-frequency continuum agreement; smaller values reduce the impact of these auxiliary constraints. Thus, the chosen values were selected as practical settings that balance the recovery of local features with the reconstruction of large-scale continuum, rather than as formally optimized hyperparameters. We summarize each term below.

\medskip
\noindent
(1) Mean Squared Error ($L_{\mathrm{MSE}}$):
We define the primary reconstruction term as the MSE between the predicted and target fluxes over all wavelength pixels,
\begin{equation}
L_{\mathrm{MSE}} = \frac{1}{N} \sum_{i=1}^{N} \left( \hat{F}_i - F_i \right)^2,
\end{equation}
where $N$ is the number of wavelength pixels and $\hat{F}_i$ and $F_i$ are the predicted and target flux at pixel $i$, respectively. We use an unweighted MSE as the baseline pixel-space term.

\medskip
\noindent
(2) Peak Region Penalty ($L_{\mathrm{peak}}$):
Neural networks can underpredict high-contrast structures and oversmooth sharp features. To reduce this effect, we construct for each spectrum a binary mask that selects pixels whose \emph{target} flux lies above the 90th percentile of that spectrum. We then compute an MSE restricted to these high-flux pixels,
\begin{equation}
L_{\mathrm{peak}} = \frac{1}{N_{\mathrm{peak}}} \sum_{i \in \mathcal{P}} \left( \hat{F}_i - F_i \right)^2,
\end{equation}
where $\mathcal{P}$ is the set of masked pixels and $N_{\mathrm{peak}} = |\mathcal{P}|$. This term increases the weight on continuum peaks and other high-flux structures.

\medskip
\noindent
(3) Fourier Domain Loss ($L_{\mathrm{fourier}}$):
To constrain the large-scale continuum shape, we include a loss in the Fourier domain. Before applying the discrete Fourier transform, we replace the first and last flux values with their nearest neighbors to reduce edge artifacts. We then compute a real-valued fast Fourier transform along the wavelength axis and compare only the lowest $N_{\mathrm{low}}$ frequency modes:
\begin{equation}
\begin{array}{rl}
L_{\mathrm{fourier}}
=& \displaystyle
\frac{1}{N_{\mathrm{low}}}
\sum_{k=1}^{N_{\mathrm{low}}}
\left[
\Delta \Re_k^{2} + \Delta \Im_k^{2}
\right], \\[6pt]
\Delta \Re_k
=& \displaystyle
\Re\!\left[\tilde{F}_{\mathrm{pred}}(k)\right]
-
\Re\!\left[\tilde{F}_{\mathrm{true}}(k)\right], \\[6pt]
\Delta \Im_k
=& \displaystyle
\Im\!\left[\tilde{F}_{\mathrm{pred}}(k)\right]
-
\Im\!\left[\tilde{F}_{\mathrm{true}}(k)\right].
\end{array}
\end{equation}
where $\tilde{F}(k)$ is the $k$th Fourier coefficient. In all experiments we set $N_{\mathrm{low}}=500$, which captures large-scale continuum variations while remaining insensitive to high-frequency noise.

\subsection{Training Strategy}

We implement EUT in PyTorch Lightning and train with data parallelism on four NVIDIA A6000 GPUs. The per-GPU batch size is 120 (effective batch size 480). We use the AdamW optimizer with an initial learning rate of $5\times10^{-5}$ and mixed-precision training to reduce memory use and increase throughput. We adopt a \texttt{ReduceLROnPlateau} schedule, decreasing the learning rate by a factor of 0.5 if the validation loss does not improve for 10 consecutive epochs.

\begin{figure*}[ht!]
\epsscale{1.}
\plotone{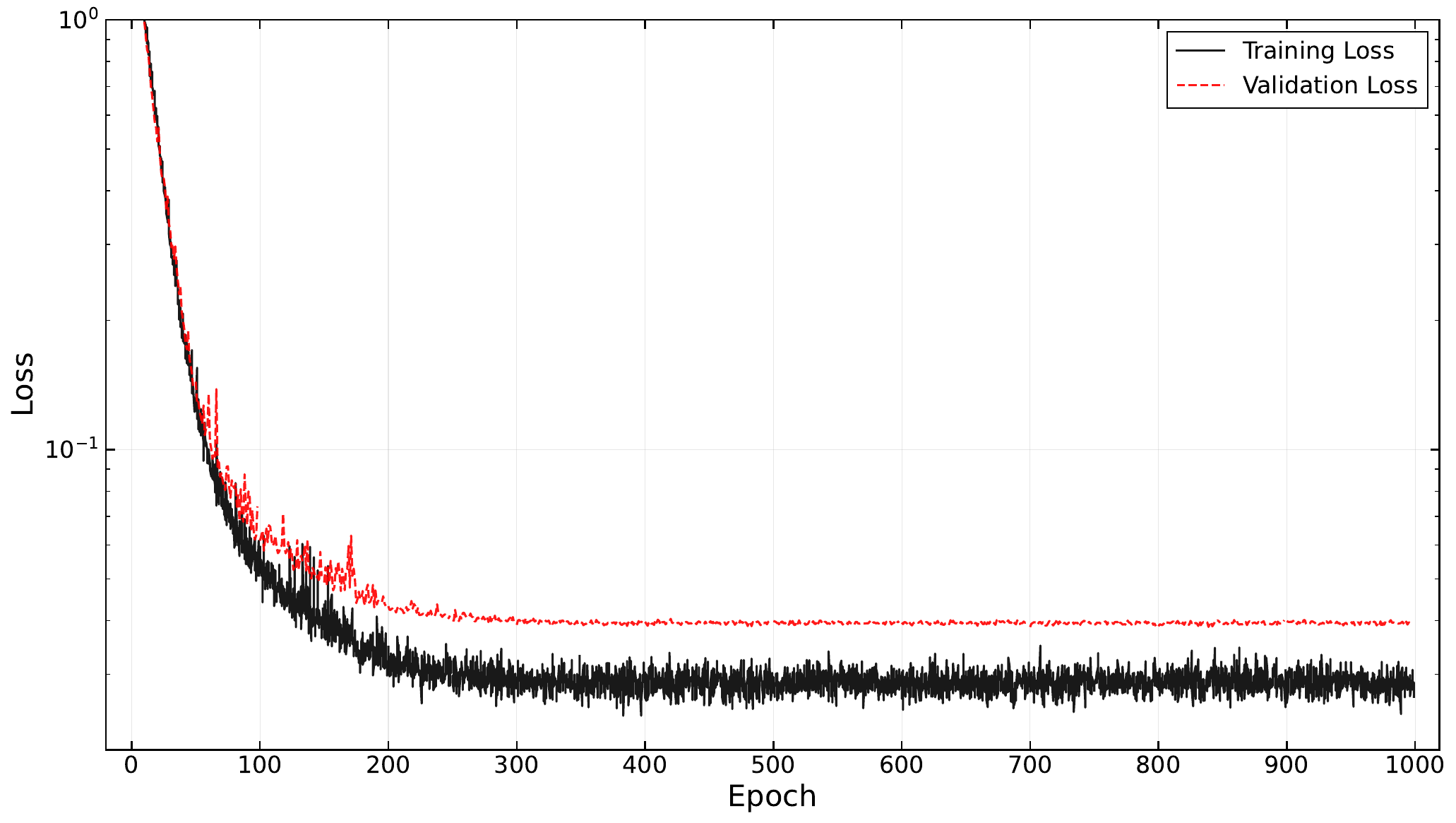}
\caption{
Training (black) and validation (red) loss as a function of epoch. The losses decrease rapidly at early epochs and then approach a plateau; the minimum validation loss occurs at epoch 974. We adopt the corresponding checkpoint for subsequent analysis.
}
\label{loss}
\end{figure*}

Training is monitored using the validation set described in Section~2. Figure~\ref{loss} shows the training and validation loss as a function of epoch. The losses decrease rapidly at early epochs and then approach a plateau. We train for up to 1000 epochs and apply early stopping with a patience of 500 epochs based on the validation loss. For the experiments presented here, early stopping is not triggered; the minimum validation loss occurs at epoch 974, and we use the corresponding checkpoint for all subsequent evaluations.

\section{Results}

We evaluate EUT using the independent test set of 10{,}000 synthetic spectra described in Section~2.3. Our analysis has three components: (1) overall denoising performance based on visual inspection and summary metrics, (2) recovery of key absorption features relevant to stellar population analysis, and (3) accuracy of mass-weighted age and metallicity recovered with full-spectrum fitting applied to the denoised spectra.

\subsection{Global Spectral Denoising Performance}

\begin{figure*}[ht!]
\epsscale{1.1}
\plotone{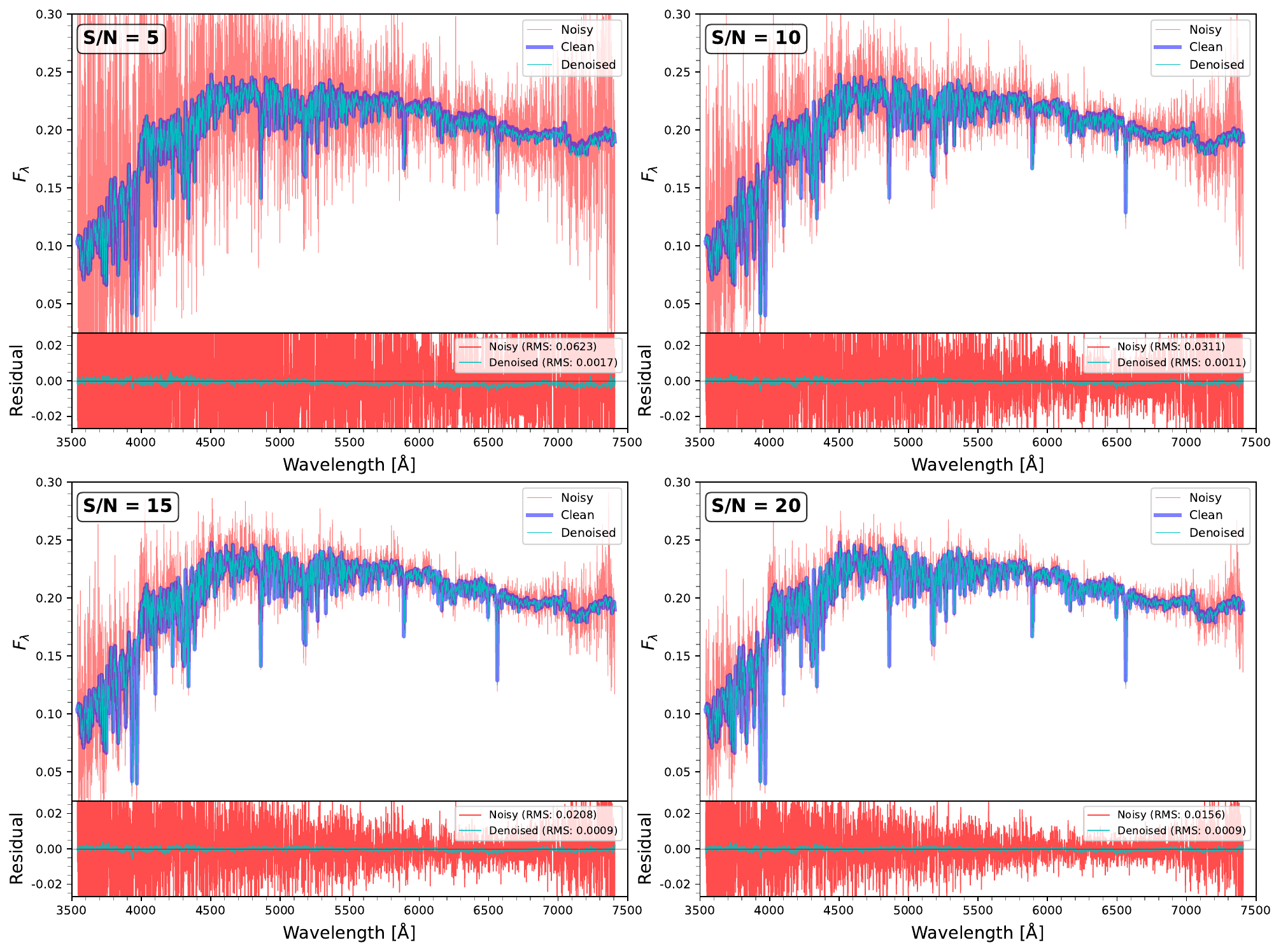}
\caption{
Example of full-spectrum denoising results for a representative synthetic spectrum at four input signal-to-noise ratios ($S/N=5, 10, 15,$ and 20). In each panel, the top plot displays the noise-free reference (blue), the noisy input (red), and the EUT-denoised output (cyan). The bottom plot shows the residuals relative to the noise-free spectrum for both the noisy input (red) and the denoised output (cyan). The legend indicates the root-mean-square (RMS) residual for each case.
}
\label{fig:full_spectrum_examples}
\end{figure*}

\begin{figure*}[ht!]
\epsscale{1.2}
\plotone{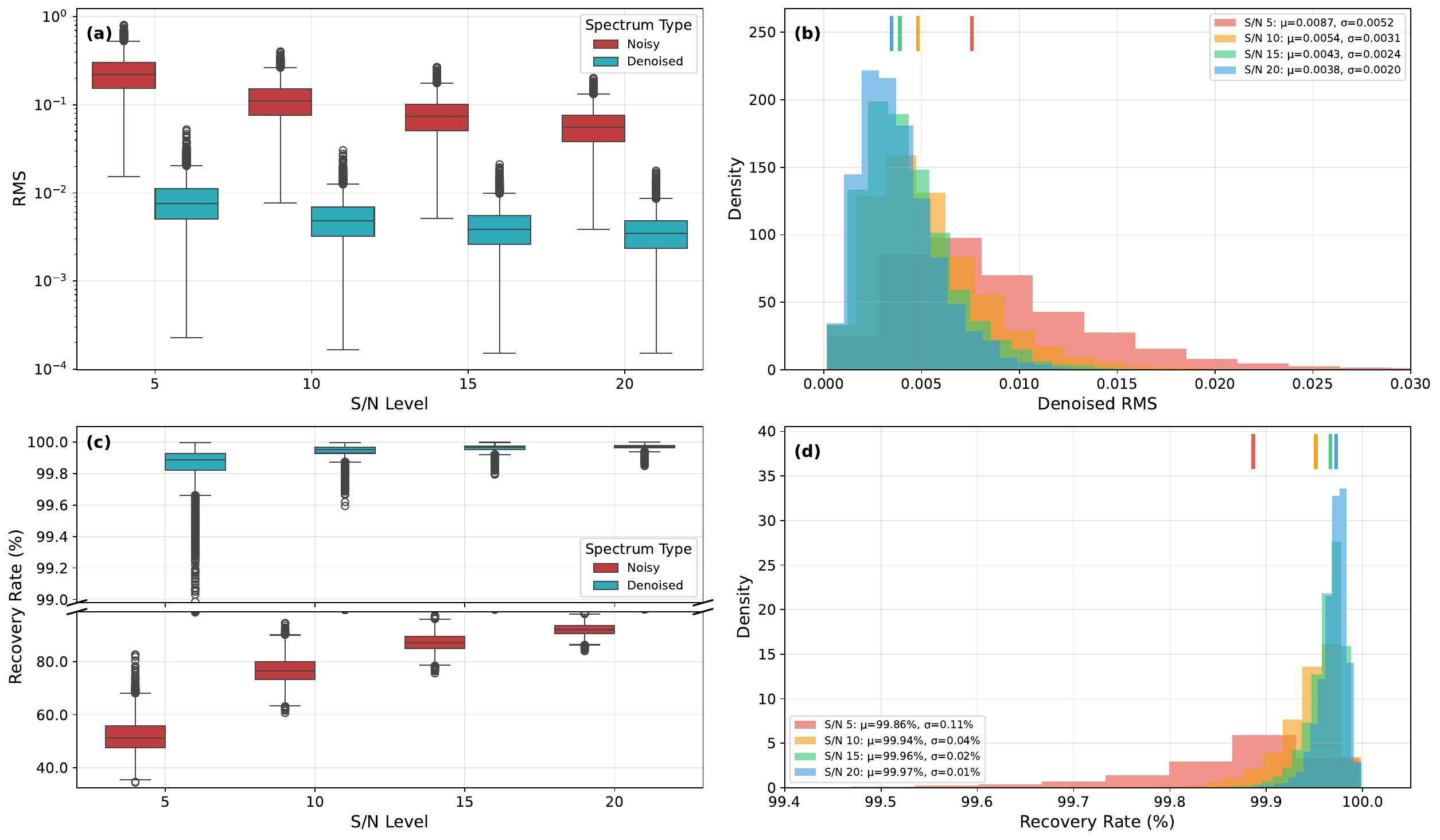}
\caption{Full-spectrum denoising statistics for 10{,}000 synthetic test spectra at input $S/N=5$, 10, 15, and 20. (a) Box-and-whisker plots of per-spectrum RMS residuals relative to the corresponding noise-free spectra, comparing noisy inputs (red) and EUT-denoised outputs (cyan). (b) Distributions of RMS residuals for the denoised outputs; vertical ticks mark medians, and the legend lists the mean ($\mu$) and standard deviation ($\sigma$) at each $S/N$. (c) Recovery rate, defined as the Pearson correlation coefficient between the noise-free and comparison spectra (Equation~\ref{eq:recovery_rate}) and expressed as a percentage; the y-axis is broken to emphasize the denoised results. (d) Distributions of recovery rates for the denoised outputs; the legend lists $\mu$ and $\sigma$. In panels (a) and (c), the box spans the interquartile range (IQR), the horizontal line marks the median, whiskers extend to 1.5$\times$IQR, and black open circles indicate outliers.}
\label{fig:global_denoising_stats}
\end{figure*}

We evaluate the ability of EUT to recover the noise-free continuum shape and suppress noise over the full wavelength range.
Figure~\ref{fig:full_spectrum_examples} shows a randomly selected synthetic spectrum at four representative input signal-to-noise ratios ($S/N = 5, 10, 15, 20$). In each case, the denoised spectra (cyan) closely follow the noise-free references (blue) across the wavelength range, even when the noisy inputs (red) exhibit strong high-frequency fluctuations.
The residuals (lower panels) show that EUT effectively reduces small-scale noise. Since this figure presents just one spectrum from the 10{,}000-spectrum test set, any localized signed offsets in an individual panel---such as the minor offset observed at the long-wavelength end in the low-$S/N$ case---should not be taken as evidence of a systematic wavelength-dependent bias in the model. Instead, the overall behavior of the entire test set is encapsulated by the population-level statistics shown in Figure~\ref{fig:global_denoising_stats}.

To quantify performance, we use the independent test set of 10{,}000 synthetic spectra (Section~2.3) and compute the root-mean-square (RMS) residual relative to the noise-free spectrum for both the noisy inputs and the denoised outputs. Figure~\ref{fig:global_denoising_stats}(a) and (b) compare the RMS distributions as a function of input $S/N$. At $S/N=5$, the median RMS decreases from 0.221 (noisy) to $7.7\times10^{-3}$ (denoised), corresponding to a 96.5\% reduction. At $S/N=20$, the median RMS decreases from $5.3\times10^{-2}$ to $3.0\times10^{-3}$ (94\% reduction). Similar improvements are obtained at intermediate $S/N$.

We also assess similarity between the noise-free ($f_{\mathrm{clean}}$) and denoised ($f_{\mathrm{denoised}}$) spectra using the Pearson correlation coefficient, reported as a percentage (``recovery rate''):

\begin{equation}
\begin{array}{rl}
\mathcal{R}
=& \displaystyle
\frac{\sum_i c_i d_i}
{\left(\sum_i c_i^2\right)^{1/2}
 \left(\sum_i d_i^2\right)^{1/2}}
\times 100, \\[6pt]
c_i
=& f_{\mathrm{clean},i}-\bar{f}_{\mathrm{clean}}, \\[4pt]
d_i
=& f_{\mathrm{denoised},i}-\bar{f}_{\mathrm{denoised}}.
\end{array}
\label{eq:recovery_rate}
\end{equation}
where $\bar{f}$ denotes the mean flux over wavelength pixels. As shown in Figure~\ref{fig:global_denoising_stats}(c), the EUT-denoised outputs achieve median recovery rates of 99.9\% at $S/N=5$, rising to 99.97\% at $S/N=15$--20, whereas the noisy spectra (red boxes) typically yield values below 90\%.
Figure~\ref{fig:global_denoising_stats}(d) further shows the full recovery-rate distributions for the denoised outputs, which directly quantify the spectrum-to-spectrum scatter and the size of any low-recovery tail. Even at $S/N=5$, the distribution is sharply peaked near 100\% ($\mu=99.86\%$, $\sigma=0.11\%$), and it becomes progressively narrower toward higher $S/N$ (e.g., $\sigma=0.04\%$ at $S/N=10$ and $\sigma=0.01\%$ at $S/N=20$).
These RMS and correlation statistics show that EUT substantially improves reconstruction fidelity for low-$S/N$ spectra in our synthetic tests.

\subsection{Recovery of Representative Absorption Lines}
\label{sec:absorption_lines}

\begin{figure*}[ht!]
\epsscale{1.1}
\plotone{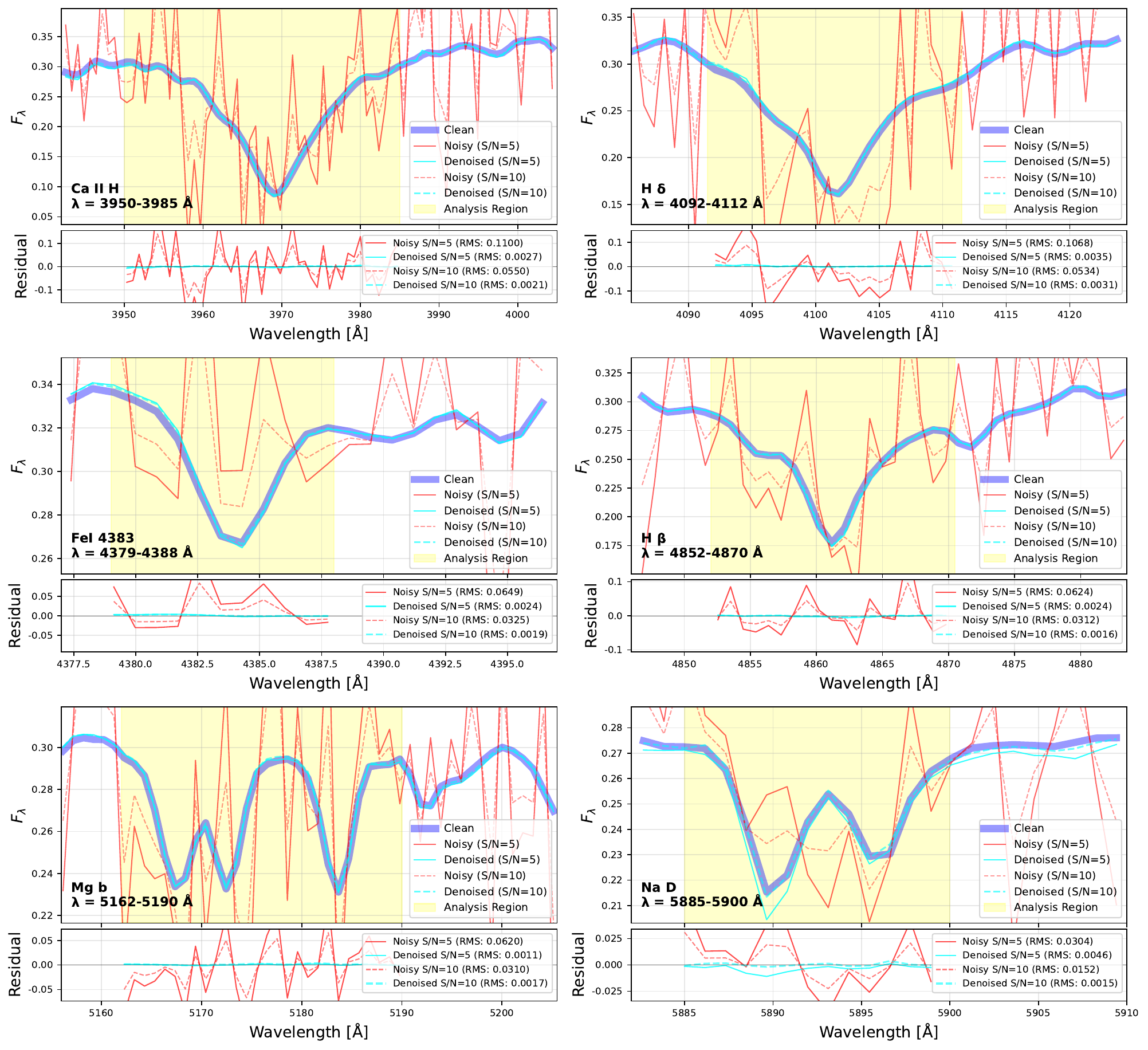}
\caption{
Example of absorption-line recovery at input $S/N=5$ and 10. Columns correspond to six diagnostic features: \ion{Ca}{2}~H, H$\delta$, Fe\,\textsc{i}\,4383, H$\beta$, Mg~b, and Na~D. In each column, the upper panel overlays the noise-free spectrum (blue), the noisy inputs (red), and the denoised outputs (cyan). Solid lines represent $S/N=5$, while dashed lines represent $S/N=10$. The shaded regions mark the fixed wavelength windows used for quantitative analysis. The lower panel shows the residuals relative to the noise-free spectrum; the legend reports the RMS residual within the analysis window.
}
\label{fig:absorption_examples}
\end{figure*}

Absorption features provide key constraints on stellar age, metallicity, and abundance ratios. We therefore examine how well EUT recovers representative absorption lines from low-$S/N$ spectra. Figure~\ref{fig:absorption_examples} shows six diagnostic features---\ion{Ca}{2}~H, H$\delta$, Fe\,\textsc{i}\,4383, H$\beta$, Mg~b, and Na~D---for a single spectrum at input $S/N=5$ and 10. In the noisy spectra (red), line profiles are strongly distorted, particularly at $S/N=5$, where line cores and wings are obscured by high-frequency fluctuations. The EUT-denoised spectra (cyan) closely follow the noise-free references (blue) and recover the overall morphology of each feature, including line depth and width, in both cases.

\begin{figure*}
\centering
\includegraphics[width=\textwidth]{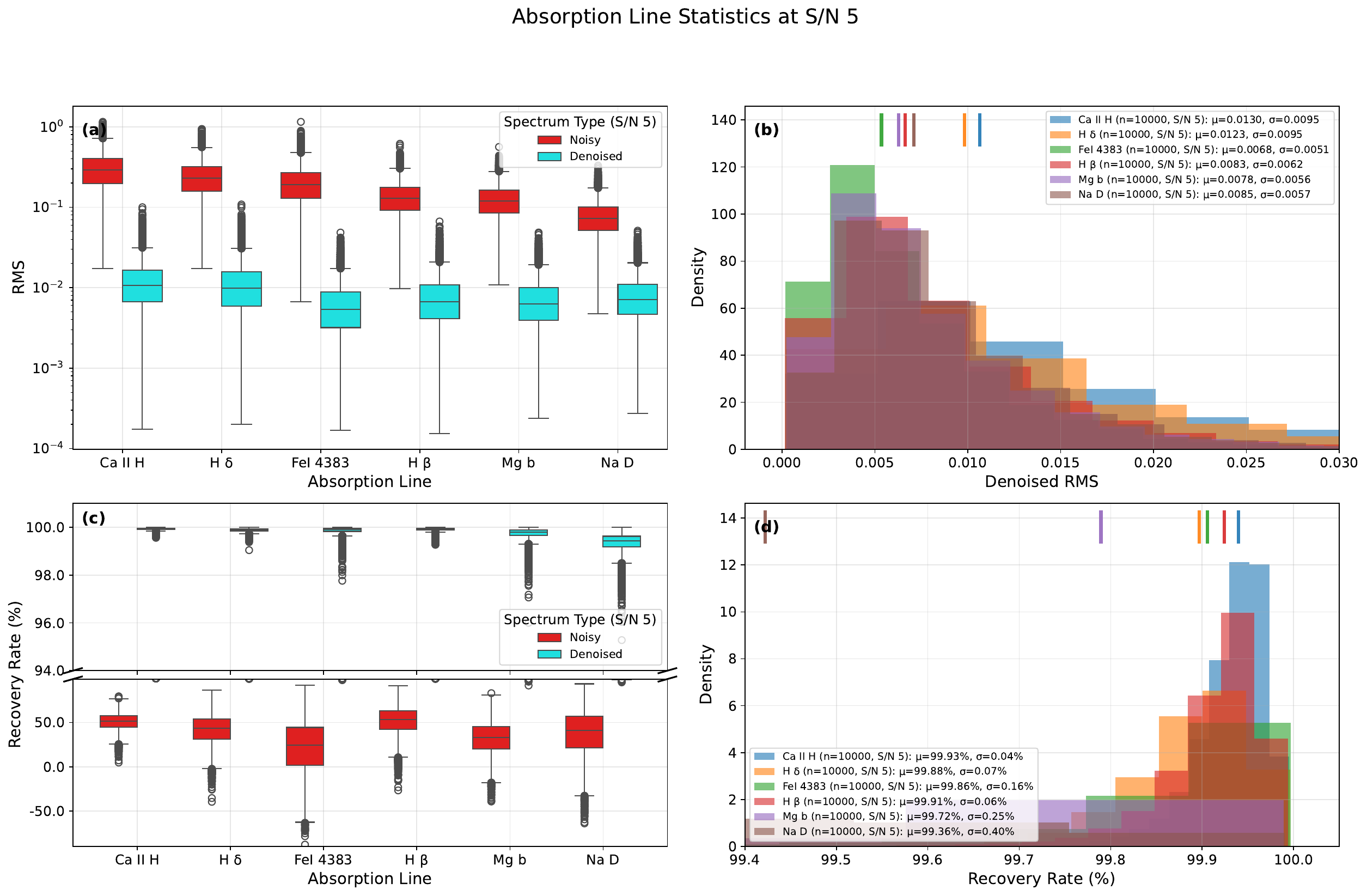}
\caption{Absorption-line denoising statistics at input $S/N=5$ for six diagnostic features (\ion{Ca}{2}~H, H$\delta$, \ion{Fe}{1}~4383, H$\beta$, Mg~b, and Na~D). Measurements use the fixed wavelength windows highlighted in Figure~\ref{fig:absorption_examples} for 10{,}000 synthetic test spectra. (a) Box-and-whisker plots of RMS residuals (relative to the noise-free spectra) within each window for noisy inputs (red) and denoised outputs (cyan), shown on a logarithmic scale. (b) Distributions of the denoised RMS residuals; vertical ticks mark medians, and the legend lists the mean ($\mu$) and standard deviation ($\sigma$) for each feature. (c) Recovery rate within each window (Pearson correlation coefficient; Equation~\ref{eq:recovery_rate}), with a broken y-axis for clarity. (d) Distributions of the denoised recovery rates; the legend lists $\mu$ and $\sigma$ (in \%). Boxplot conventions (including black open-circle outliers) follow Figure~\ref{fig:global_denoising_stats}.}
\label{fig:absorption_stats}
\end{figure*}

To quantify this behavior, we measure the RMS residual within fixed wavelength windows around each feature for 10{,}000 test spectra at $S/N=5$. Figure~\ref{fig:absorption_stats} (a) and (b) shows the distributions of RMS residuals before and after denoising for the absorption line features. The residual RMS decreases for all lines after denoising. Fe\,\textsc{i}\,4383, the narrowest feature in our set, shows the largest improvement: the RMS residual is reduced by $96.5 \pm 2.3\%$, from $0.208 \pm 0.109$ in the noisy spectra to $0.007 \pm 0.005$ after denoising. \ion{Ca}{2}~H, located in the blue where the noise level is higher, shows an RMS reduction of $95.8 \pm 2.1\%$, with the residual RMS decreasing from 0.314 to 0.013. The Balmer lines H$\delta$ and H$\beta$ show RMS reductions of $95.1 \pm 2.8\%$ and $93.9 \pm 3.5\%$, respectively. The Mg~b feature is recovered with an RMS reduction of $93.6 \pm 3.9\%$. The Na~D doublet shows the smallest, but still substantial, improvement of $88.4 \pm 6.9\%$. This lower value likely reflects that Na~D is weak in a subset of the synthetic spectra, making the residuals more sensitive to noise. Across these features, the RMS reduction exceeds 88\% for all lines, indicating that EUT improves recovery of absorption-line structure at low $S/N$ in our synthetic tests.
Because a small RMS does not by itself guarantee that the \emph{shape} of a line profile is recovered, we also evaluate the recovery rate (Equation~\ref{eq:recovery_rate}) within each line window. Figure~\ref{fig:absorption_stats}(c) shows that, at $S/N=5$, the noisy line windows often yield low (and in some cases negative) correlations, suggesting profiles dominated by noise fluctuations. In contrast, the denoised outputs are closely aligned with 100\% recovery for all six features. The corresponding distributions in Figure~\ref{fig:absorption_stats}(d) quantify the feature-to-feature scatter: \ion{Ca}{2}~H and H$\beta$ reach $\mu\simeq99.9\%$ with $\sigma\lesssim0.06\%$, while Mg~b and Na~D show broader recovery-rate distributions ($\mu=99.72\%$, $\sigma=0.25\%$ for Mg~b; $\mu=99.36\%$, $\sigma=0.40\%$ for Na~D). The broader scatter for Na~D is consistent with this feature being relatively weak in a subset of spectra, which makes the correlation metric more sensitive to residual fluctuations. Nevertheless, even the most challenging feature maintains a high mean recovery rate, supporting that EUT generally preserves line-profile morphology within the analysis windows.

\section{Impact on Stellar Population Recovery}
\label{sec:impact_spp}

\begin{figure*}
\centering
\includegraphics[width=\textwidth]{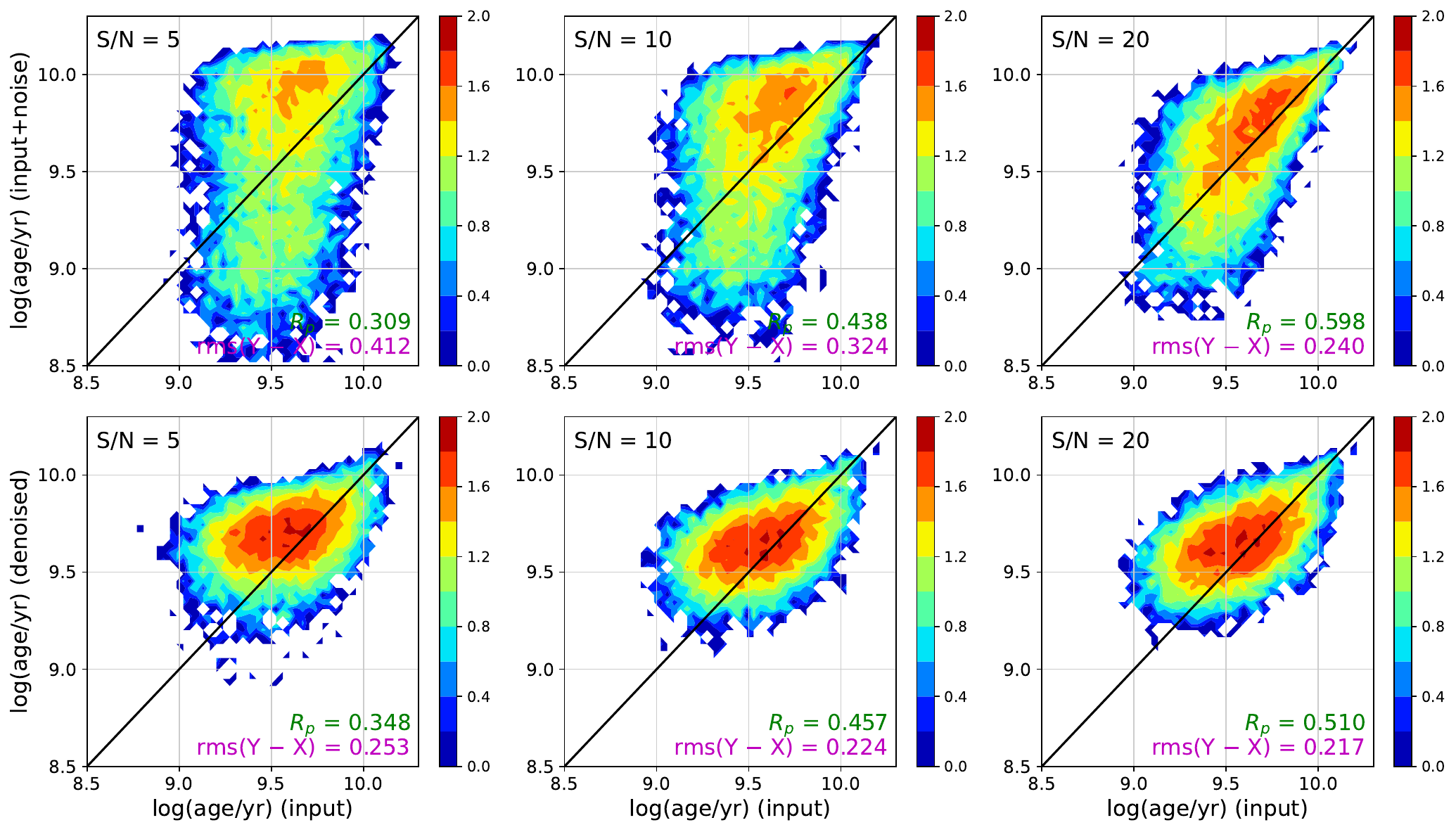}
\caption{
Recovery of mass-weighted age from noisy and EUT-denoised spectra. Each panel compares the input age [$\log(\mathrm{age/yr})$; $x$-axis] with the value recovered by \textsc{pPXF} ($y$-axis) for $10^{4}$ synthetic spectra. Colors show the logarithmic number density, and the black line indicates the one-to-one relation. The top row shows results for the noisy spectra, and the bottom row shows results for the corresponding denoised spectra. Columns correspond to $S/N=5$, 10, and 20 (left to right). Each panel reports the Pearson correlation coefficient, $R_p$, and the rms difference, $\mathrm{rms}(Y-X)$, in dex. Denoising reduces the scatter around the one-to-one relation, particularly at $S/N \leq 10$.
}
\label{fig:age_recovery}
\end{figure*}

\begin{figure*}
\centering
\includegraphics[width=\textwidth]{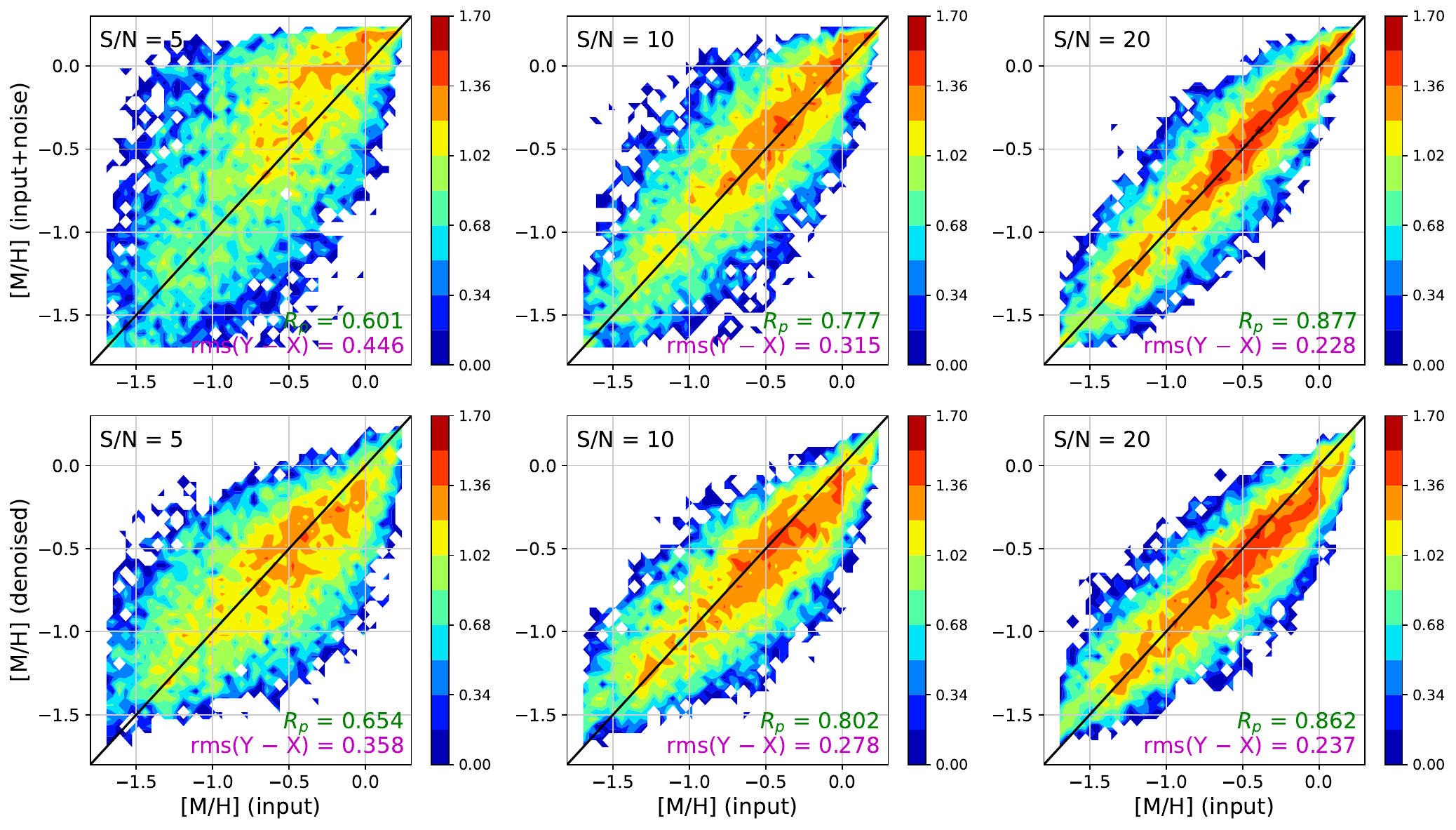}
\caption{
Same as Figure~\ref{fig:age_recovery}, but for mass-weighted global metallicity, $[\mathrm{M/H}]$. Each panel compares the input metallicity ($x$-axis) with the value recovered by \textsc{pPXF} ($y$-axis) from the noisy (top row) and denoised (bottom row) spectra. Colors show the logarithmic number density, and the diagonal line indicates the one-to-one relation. Columns correspond to $S/N=5$, 10, and 20 (left to right). Each panel reports the Pearson correlation coefficient, $R_p$, and the rms difference, $\mathrm{rms}(Y-X)$, in dex.
}
\label{fig:met_recovery}
\end{figure*}

To quantify how denoising affects downstream stellar-population inference, we fit both noisy and denoised spectra with \textsc{pPXF} and compare the recovered mass-weighted age and metallicity to the known input values. Figures~\ref{fig:age_recovery} and \ref{fig:met_recovery} show recovered versus input parameters for $10^{4}$ spectra at $S/N=5$, 10, and 20. Each panel shows the logarithmic number density, the one-to-one relation, the Pearson correlation coefficient ($R_p$), and the rms difference $\mathrm{rms}(Y-X)$.

We fit each spectrum with \textsc{pPXF} \citep{Cappellari2017} to recover mass-weighted stellar population parameters. We perform noise-weighted fits using an input noise (or equivalently, $S/N$) vector as a function of wavelength, ensuring that higher-$S/N$ wavelength regions receive higher weight while low-$S/N$ portions contribute less to the solution, following the practical approach of \citet{Lee2023}.

To ensure a consistent analysis between the noisy inputs and the denoised outputs (which lack intrinsic propagated error variance), we determine the weighting through an empirical two-pass procedure. First, we run \textsc{pPXF} assuming a constant noise vector (uniform weights) to obtain an initial best-fitting model. We then estimate a wavelength-dependent effective noise based on the residual fluctuations between the input spectrum and this initial model. Finally, we rerun \textsc{pPXF} using the derived noise vector to obtain the final fit and stellar population parameters. This same procedure is applied consistently to both the noisy and denoised spectra.

For mass-weighted age (upper panels of Figure~\ref{fig:age_recovery}), the noisy spectra yield broad distributions with weak correlation at low $S/N$. At $S/N=5$, the correlation between input and recovered age is modest ($R_{p} \simeq 0.31$) and the scatter is large ($\mathrm{rms}(Y-X) \simeq 0.41$~dex).
Although EUT helps to minimize random variations, a systematic bias remains at the extremes of the age distribution, especially in the low-$S/N$ range. As illustrated in Figure~\ref{fig:age_recovery}, the youngest stellar populations ($\sim 10^{8.5}$\,yr) are still systematically estimated to be older than their input values, while the oldest populations show a slight bias toward younger ages.
This ``regression to the mean'' effect is likely driven by the increased noise in the blue spectral region, which contains crucial age diagnostics. When these features are degraded by noise, the recovery becomes less constrained and tends to bias toward the average age of the sample. After denoising, the distribution becomes more concentrated around the one-to-one relation, but substantial scatter remains at $S/N=5$, and the systematic overestimation at the youngest ages is not fully removed; the rms decreases to $\sim 0.25$~dex, corresponding to a $\approx 40\%$ reduction, while $R_p$ changes only slightly. At $S/N=10$, the rms improves from $\sim 0.32$~dex to $\sim 0.22$~dex ($\approx 30\%$ reduction), again with a small change in $R_p$. At $S/N=20$, where the noisy spectra already yield relatively small scatter ($\mathrm{rms} \simeq 0.24$~dex), denoising reduces the rms to $\sim 0.22$~dex. At intermediate and high $S/N$, $R_p$ changes little after denoising. This indicates that EUT mainly reduces scatter about the one-to-one relation rather than improving the rank ordering; residual age--metallicity degeneracy in the \textsc{pPXF} fits likely limits further improvement.

For mass-weighted metallicity (Figure~\ref{fig:met_recovery}), the trends are similar but less pronounced. At $S/N=5$, denoising increases the correlation from $R_{p} \simeq 0.60$ to $R_{p} \simeq 0.65$ and reduces the scatter from $\mathrm{rms}(Y-X) \simeq 0.45$~dex to $\sim 0.36$~dex ($\approx 20\%$ reduction). At $S/N=10$, the rms decreases from $\sim 0.32$~dex to $\sim 0.28$~dex, with a small increase in $R_p$ (from $\simeq 0.77$ to $\simeq 0.80$). At $S/N=20$, metallicity recovery from the noisy spectra is already robust ($R_{p}\simeq0.87$, ${\rm rms}\simeq0.23$~dex), and applying denoising changes these summary metrics only modestly.

Overall, EUT denoising improves the precision of stellar population parameters recovered from full-spectrum fitting in the low- and intermediate-$S/N$ regimes, while changes at $S/N=20$ are modest. The largest gains are observed for mass-weighted age at $S/N \lesssim 10$, where the rms decreases by $\sim 30$--40\%, with smaller but measurable improvements for metallicity at the lowest $S/N$.

\section{Summary and Conclusions}
\label{sec:summary}

To address the limitations imposed by low-$S/N$ spaxels in modern IFU surveys without relying on spatial binning, we investigate whether deep-learning-based denoising technique can improve low-$S/N$ spectra while preserving the original spatial resolution.
Given that the noise-free spectrum and true stellar population parameters are not available for real galaxies, we perform controlled synthetic experiments where both are specified. We present the EUT, a 1D hybrid architecture that combines a convolutional encoder--decoder with a Transformer bottleneck. The model is trained using a composite loss function that constrains pixel-space residuals, high-flux regions, and the overall continuum shape.

We train EUT on synthetic galaxy spectra constructed from MILES SSP models, as outlined by \citet{Lee2023}, with wavelength-dependent noise injected on the fly to mimic IFU-like observations across $S/N \simeq 5$--20 (as defined in the specified continuum window). This framework enables a direct comparison between denoised outputs and their corresponding noise-free references.

On the full-spectrum level, the application of EUT improves agreement with the noise-free spectra in our test set (Figures~\ref{fig:full_spectrum_examples} and \ref{fig:global_denoising_stats}). Using the per-spectrum RMS residual, evaluated relative to the noise-free spectrum across the entire wavelength range, the median RMS decreases at $S/N=5$ from 0.221 for the noisy inputs to $7.7\times10^{-3}$ for the denoised outputs (a 96.5\% reduction). At $S/N=20$, the median RMS decreases from $5.3\times10^{-2}$ to $3.0\times10^{-3}$ (a 94\% reduction). The Pearson correlation coefficient between the noise-free and denoised spectra, denoted as the recovery rate (Equation~\ref{eq:recovery_rate}), exhibits a median of 99.9\% at $S/N=5$ and reaches 99.97\% at $S/N=15$--20, whereas the noisy spectra yield substantially lower values.

Within fixed wavelength windows surrounding representative absorption features (\ion{Ca}{2}~H, H$\delta$, Fe\,\textsc{i}\,4383, H$\beta$, Mg~b, and Na~D), EUT also effectively reduces residual structure at low $S/N$ (Figures~\ref{fig:absorption_examples} and \ref{fig:absorption_stats}). At $S/N=5$, the residual RMS within these windows decreases by at least 88\% for all examined lines, with the most significant reductions approaching $\sim 96\%$ for the narrowest features. These results indicate that EUT successfully suppresses high-frequency noise while maintaining the morphology of absorption lines within the analysis windows used for quantitative evaluation.

We evaluate the influence of denoising on subsequent stellar population measurements by employing full-spectrum fitting with \textsc{pPXF} on both noisy and denoised spectra, using the same fitting setup (Figures~\ref{fig:age_recovery} and \ref{fig:met_recovery}). Regarding mass-weighted age, the rms difference between the recovered and true values decreases from $\simeq 0.41$ to $\simeq 0.25$~dex at $S/N=5$ and from $\simeq 0.32$ to $\simeq 0.22$~dex at $S/N=10$, while remaining comparable ($\simeq 0.22$~dex) at $S/N=20$. For mass-weighted $[\mathrm{M/H}]$, the rms decreases from $\simeq 0.45$ to $\simeq 0.36$~dex at $S/N=5$ and from $\simeq 0.32$ to $\simeq 0.28$~dex at $S/N=10$, with only minor changes at $S/N=20$. In this synthetic setting, denoising improves the precision of recovered parameters at low and intermediate $S/N$; at $S/N=20$, the recovered parameters are largely comparable between the noisy and denoised spectra.

We acknowledge several limitations in the current framework that motivate our future work. First, the current model is exclusively trained and validated on synthetic spectra.
Consequently, its ability to generalize to real IFU observations remains uncertain, as these observations may involve additional systematic errors (e.g., flux-calibration errors, sky-subtraction residuals, wavelength-dependent covariance, imperfect masking, and instrumental line-spread-function variations) which are not represented in the training dataset. Therefore, we treat the present results as a proof of concept in a controlled setting rather than a ready-to-use tool for immediate application to survey data.

A primary goal of forthcoming work is to develop a denoiser that is applicable to real IFU spectra. To achieve this, we will train and evaluate models using observed data products, leveraging high-$S/N$ spectra as empirical references and generating controlled low-$S/N$ realizations through noise injection and downsampling strategies that preserve the survey's noise characteristics. We will also explore domain-adaptation approaches---including fine-tuning on observed spectra and hybrid training that combines synthetic and observed examples---to reduce the domain gap between simulations and empirical data. In parallel, we will expand the training library to incorporate additional astrophysical and observational realism (e.g., dust attenuation, emission-line contamination, and survey-specific calibration artifacts) and will assess robustness across multiple IFU data sets with different instrumental characteristics.

Second, we aim to refine the model's performance and broaden its scientific scope.
We will specifically address the systematic bias observed in age recovery for young stellar populations (Section 5) by implementing wavelength-dependent loss weighting to emphasize the age-sensitive blue spectral region or by adopting curriculum learning for the low-$S/N$ regime.
Building on this improved fidelity, future extensions will move beyond the current mass-weighted parameters to investigate the recovery of detailed multi-component star-formation and chemical-enrichment histories.
Finally, we will incorporate uncertainty quantification to ensure that denoising and downstream fitting can be propagated into statistically interpretable parameter errors. At a practical level, we plan to estimate predictive uncertainties in the denoised spectra using stochastic or ensemble-based neural-network predictions. These uncertainties will then be propagated into the recovered stellar-population parameters through repeated downstream fitting.

A complementary future approach is to employ neural networks to directly infer stellar population parameters from noisy spectra, bypassing the initial step of reconstructing a denoised spectrum before applying \textsc{pPXF}. This end-to-end method tackles a related but distinct issue, as it substitutes the fitting step with a learned mapping from spectra to physical parameters. It could offer computational benefits during application by eliminating the iterative fitting process; however, it would no longer isolate the impact of spectral denoising on downstream fitting performance. The current study is thus designed to assess whether enhancing spectral quality alone results in more reliable parameter recovery within a standard fitting framework, while ensuring direct comparability with \citet{Lee2023}. Investigating CNN–Transformer models that directly predict stellar population parameters, thereby approximating part of the role currently fulfilled by \textsc{pPXF}, will be a crucial direction for future research.

Consequently, the current proof-of-concept results indicate that hybrid CNN--Transformer denoisers such as EUT can improve low-$S/N$ spectra in controlled tests without the need for spatial binning. This outcome motivates a subsequent phase focused on training and validating models that are ready for application to real IFU survey data.

\begin{acknowledgments}
This work was supported by the National Research Foundation of Korea through grants, NRF-2022R1C1C2005539 (S.K.),NRF-2022R1A2C1004025 (J.H.L.), and NRF-2022R1A2C1007721 (S.C.R.).
\end{acknowledgments}

\bibliography{sample7}{}
\bibliographystyle{aasjournalv7}

\end{document}